\documentclass[iop]{emulateapj}

\begin{document}

\title{TIMING ANALYSIS OF THE PERIODIC RADIO AND OPTICAL BRIGHTNESS VARIATIONS OF THE ULTRACOOL DWARF, TVLM 513-46546}

\author{A. Wolszczan\altaffilmark{1,2} \& M. Route\altaffilmark{1,2,3}}

\altaffiltext{1}{Department of Astronomy and Astrophysics, the Pennsylvania State University, 525 Davey Laboratory, University Park, PA 16802, alex@astro.psu.edu}

\altaffiltext{2}{Center for Exoplanets and Habitable Worlds, the Pennsylvania State University, 525 Davey Laboratory, University Park, PA 16802}

\altaffiltext{3}{Present address: Northrop Grumman Electronic Systems, 6120 Longbow Drive, Boulder, CO 80301, matthew.route@ngc.com}

\slugcomment{Accepted for publication in ApJ; April 17, 2014}

\begin{abstract}
We describe the arrival time measurements and timing modeling of the periodic radio flares and optical brightness variations of the M9 ultracool dwarf, TVLM 513-46546. We confirm the stability of the observed period and determine its best-fit value to be 7054.468$\pm$0.007 s over the last 7 years, based on both the new and archival radio observations and the archival optical data. The period, when measured separately for the radio flare and the optical periodicities, is the same to within $\pm$0.02 s. We show that the radio flares are out of phase with respect to the optical brightness maxima by 0.41$\pm$0.02 of the period. Our analysis also reveals that, on shorter timescales, the period varies with the amplitude of $\pm$1-2 s about its long-term average and that these variations are correlated between the radio and the optical wavelengths. These results provide further evidence that TVLM 513-46546 is equipped with a stable, approximately dipolar magnetic field which powers the activity of the star observed over a wide wavelength range, and that the active area has been maintaining its identity and positional stability over no less than 7 years. A stepwise decline of the apparent radio flaring period of TVLM 513-46546 deduced from timing observations with the Arecibo radio telescope in late 2012 and early 2013 suggests that this effect may be the manifestation of differential rotation of the star.

\end{abstract}

\keywords{brown dwarfs - radio continuum: stars - stars: activity - stars: low-mass - stars: magnetic field - stars: rotation}
 
\section{Introduction}
Magnetic fields and rotation play an essential role in determining the internal structure and evolution of fully convective, very low mass stars and brown dwarfs (e.g. Mohanty et al. 2006; Browning 2008; McLean et al. 2012). However, despite the significant progress in our understanding of the magnetic properties of stars at the bottom of the main sequence, it remains unclear, how does the magnetic dynamo operate in these objects, and how is the magnetic energy released and conducted through their largely neutral atmospheres (Mohanty 2002; Berger et al. 2010; Reiners \& Basri 2010). In this context, radio detections of brown dwarfs and very low-mass stars cooler than the M7 spectral type, commonly referred to as the ultracool dwarfs (UCDs; Kirkpatrick et al. 1997), are important, because it is the radio emission from the UCDs, especially from the coolest ones, that provides the most direct means to diagnose their magnetic fields, trace the field geometry, and test its coupling to stellar atmospheres. Observations of circularly polarized, periodic radio emission from some UCDs, when interpreted in terms of the electron cyclotron maser mechanism (ECM, Treumann 2006, and references therein), indicate the presence of large-scale magnetic fields in these objects (Hallinan et al. 2006, 2008; McLean et al. 2011). The fields appear to have a strong, one to a few kG, dipole component (Hallinan et al. 2008), but more complicated geometries are also possible (Berger et al. 2009).

Radio emitting UCDs include examples of five periodically varying sources, three of which, TVLM 513-46546 (Hallinan et al. 2006), 2MASSW J0746425+200032 (Berger et al. 2009), and LSR J1835+3259 (Hallinan et al. 2008), emit short duration, periodic outbursts that make them vaguely ``pulsar-like", because of this behavior. The other two, 2M J00361617+1821104 (Hallinan et al. 2008), and 2M J13142039+130011 (McLean et al. 2011) exhibit much slower, periodic variations, which, at least in the case of 2M J00361617+1821104, can be interpreted as longer duration pulses or flares. Each of the first four objects has nearly identical periodicity in its radio and optical components (Lane et al. 2007; Littlefair et al. 2008; Berger et al. 2009; Harding et al. 2013). Long term, precise monitoring of the stability of such periodicities provides a direct way to determine timescales for the corresponding magnetic field stability of the UCDs, which is yet another critical constraint that needs to be established in the quest to understand the physics of these objects (e.g. Osten et al. 2006; McLean et al. 2011; Harding et al. 2013). 

We have conducted a series of observations of radio flares from one of these dwarfs, TVLM 513-46546 (hereafter TVLM 513), to investigate its periodic behavior with the aid of the timing method. We have been guided by the idea that timing measurements of periodically flaring UCDs can be used to study their physics, very much like the pulsar timing has been used as a tool to study a variety of phenomena in physics and astrophysics, including the neutron stars themselves (e.g. Blandford et al. 1992). In principle, this approach should provide a higher precision alternative to the methods of period stability analysis of TVLM 513 and other UCDs employed in the past (e.g. Doyle et al. 2010; Harding et al. 2013). We have also examined a selection of the published radio and optical measurements of TVLM 513 (Berger et al. 2008; Doyle et al. 2010; Harding et al. 2013) in an effort to phase-connect the flares from this star detected in the past with our more recent observations. 

In this paper, we report the initial results of the timing monitoring program of TVLM 513 with the Arecibo radio telescope. Also, by combining our measurements with the published data, we reassess the stability of the flaring period and the nature of the observed activity of this UCD. Observations, data analysis and timing models of the radio and the optical brightness variations of the star are described in Sections 2 and 3. In Section 4, we discuss the timing results of radio flares and the periodic optical variability of TVLM 513 since 2006, and suggest an explanation of its timing behavior over $\sim$200 days in 2012 and 2013. Our conclusions are presented in Section 5.

\section{Timing observations of TVLM 513}
TVLM 513 is an M9, $\sim$0.08 M$_\odot$ UCD, which exists just at the brown dwarf boundary (Hallinan et al. 2006). With its short, $\sim$1.96-hr rotation period and the activity that extends from GHz radio frequencies, all the way to X-rays (Hallinan et al. 2006; Berger et al. 2008), it has been arguably one of the best studied UCDs. Also, as stated above, it belongs to the small group of periodically radio flaring dwarfs, with the flaring periods tied to stellar rotation. Another factor that has contributed to our selection of this object was the persistence of its flaring generated by beamed ECM emission and the high-inclination geometry of the star's magnetic field (Hallinan et al. 2006). This is important given the tracking limitations of the Arecibo telescope, which generally restrict observations of a single source to $\le$2.5 hr day$^{-1}$.

\begin{figure}
\centering
\includegraphics[angle=0,width=0.99\textwidth]{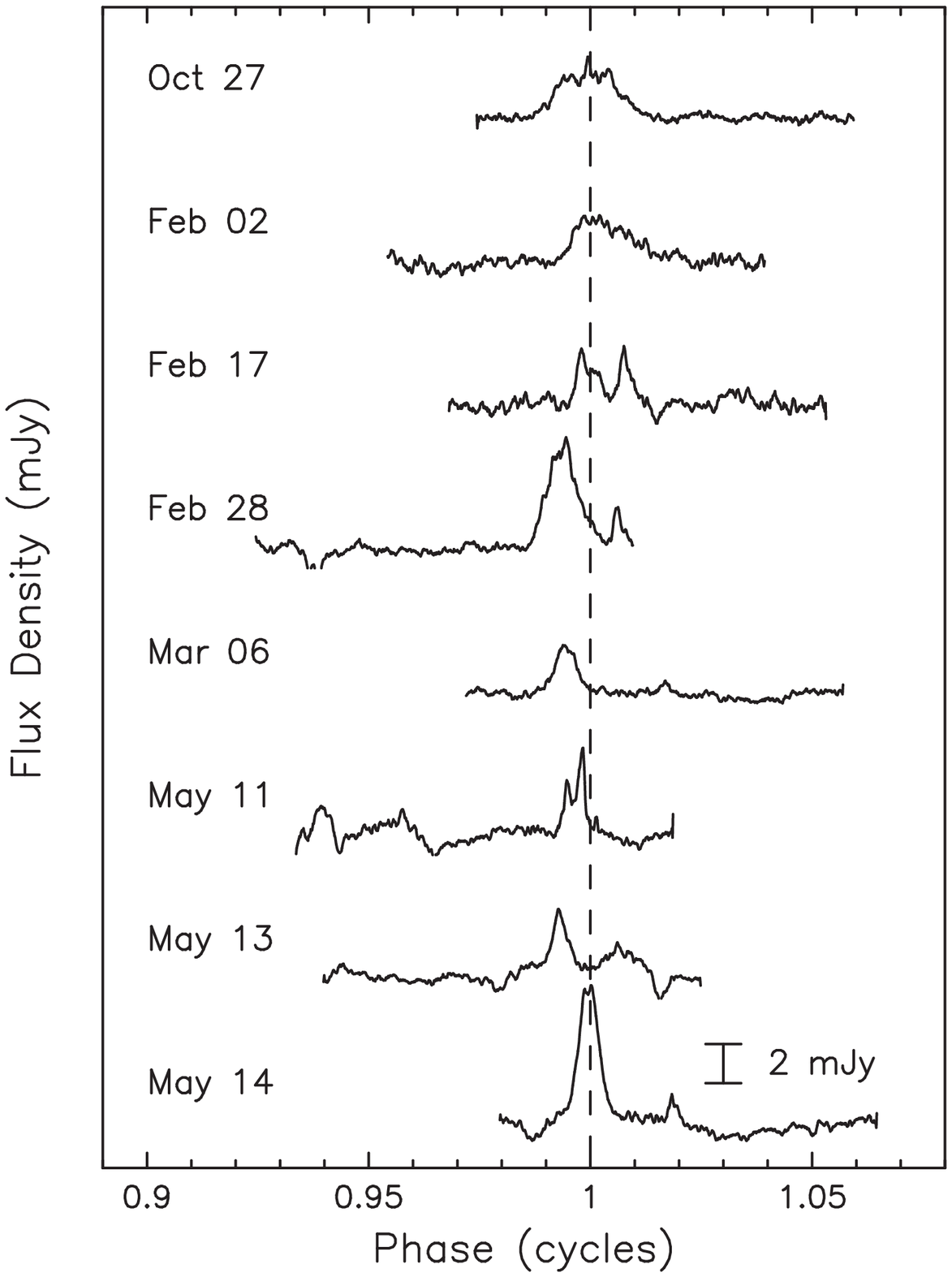}
\caption{\small Examples of circularly polarized, bandpass-averaged flares from TVLM 513 observed at 4.85 GHz with the Arecibo radio telescope in 2012 and 2013 and phased up according to the best-fit timing model discussed in the text. The vertical, dashed line marks the phase of flare arrival times predicted by the model. For display purposes, the original flare time resolution of 0.9 s has been smoothed to 4.5 s.}
\label{fig1}
\end{figure}

We have observed TVLM 513 from late 2012 October until early 2013 May, with the goal to obtain a phase-connected timing solution of the detected flares. The details of the observing setup can be found in Route \& Wolszczan (2012, 2013). Briefly, we have used the 5 GHz receiver and the broadband, fast-sampled Mock spectrometer, which covers a $\sim$1 GHz bandpass in seven, 172 MHz, individually tunable blocks. As the typically high brightness temperature of the circularly polarized bursts observed from TVLM 513 at gigahertz frequencies is best explained by the ECM mechanism (Hallinan et al. 2006, 2007, 2008, 2009; Phan-Bao et al. 2007; Berger et al. 2009), we have recorded the four Stokes parameters of the received signal. The first sample of each observation was time-tagged with the observatory's hydrogen maser clock. Given the 0.9 s sampling, the practical, one-sigma, broadband sensitivity in Stokes V varied from 0.2 to 0.4 mJy depending on the interference environment. We have also included a timing measurement of our test observation made at Arecibo with the Wideband Arecibo Pulsar Processor (WAPP) on 2008 December 29. This observation also covered a $\sim$1 GHz bandpass centered at 4.85 GHz. Typical Arecibo detections of flares from the UCDs in the form of broadband, dynamic spectra are presented in Route \& Wolszczan (2013). Examples of high signal-to-noise, bandpass-averaged flares from TVLM 513 discussed here are shown in Fig. \ref{fig1}.

Flare arrival times were measured by cross-correlating the time-tagged flare profiles with a high signal-to-noise template obtained by phasing up and adding together the strongest of the observed flares. The 1$\sigma$ errors of these measurements ranged from $\sim$0.4 s for the strongest (4-6 mJy) flares to $\sim$1.4 s for the weakest (1-2 mJy) ones. The expected uncertainty of a flare time of arrival (TOA) measurement, $\sigma_{TOA}$, can be computed from the ratio of the flare width, $W$, to the signal-to-noise, $S/N$, of the flare profile (Lorimer \& Kramer 2005). Given the sensitivity of the Arecibo 5 GHz receiving system, and the above observing parameters for TVLM 513, one obtains $\sigma_{TOA}\sim$0.4 s, assuming the peak flux density and the duty cycle values for the flare to be $\sim$4 mJy and $\sim$0.025, respectively (duty cycle is defined as $W/P$, where $P$ is the flare period, and $W\sim$180 s). This estimate is in an excellent agreement with the actual errors of the cross-correlation TOA measurements. 

In order to estimate other possible contributions to these errors, such as the obvious morphological variability of the flares (Fig. \ref{fig1}), or any short-term variations of their apparent period, we have computed the residuals from fitting the flaring period to short, 1-7 day sequences of the TOAs using the timing modeling method described in the next section. Keeping these time sequences short ensured that any long-term TOA variations that were not part of the model would not affect the behavior of the post-fit residuals. The rms residuals calculated this way ranged from 10 s to 30 s, which demonstrates that the net TOA precision of our measurements was dominated by effects other than the instrumental errors, with the flare profile variability being the most likely candidate.

We have also reprocessed the Arecibo measurements made in 2009 June (Antonova at al. 2010), and visually estimated the TOAs of the peaks of TVLM 513 flares detected with the Very Large Array (VLA) in 2006 May (Hallinan et al. 2007), and April and June 2007 (Berger et al. 2008; Doyle et al. 2010) using the published plots of flare profiles and the associated UT time scales. The corresponding uncertainties of these estimates, computed by means of the period fitting method as explained above, were approximately equal to 10 s, 30 s, 70 s, and 30 s, respectively. These values are similar to the errors computed above for the more recent Arecibo data and they show once again that TOA uncertainties of the TVLM 513 radio flares are dominated by intrinsic effects. 

Finally, we have made similar estimates for the brightness maxima of the TVLM 513 light curves measured at I-band and at the Sloan $i^{'}$-band at several epochs between 2006 and 2011 (Harding et al. 2013), and for the H$\alpha$ maxima recorded in 2007 April (Berger et al. 2008). In these cases, we have visually measured the approximate TOAs of the centroids of the published light curves, defined as mid-points between the two consecutive minima of the approximately sinusoidal brightness variation. The TOAs were determined by interpolating between the whole hour ticks of the UT-calibrated time scales provided with the published light curves. The 300-350 s uncertainties of these measurements, computed as described above, are large, which is not surprising, given the approximate nature of the TOA measurements and the width of the TVLM 513 light curve. Assuming that the $S/N$ of the radio flares and the optical brightness variations was approximately the same, and taking the duty cycle of the sine wave-like optical variations to be $\sim$0.8, one obtains $\sigma_{TOA}\sim$200 s as the expected TOA uncertainty for the optical brightness variations. This is significantly but not dramatically better than the errors computed for our measurements and it shows, not unexpectedly, that the high duty cycle of the light curve is the main limiting factor of the timing precision of TVLM 513 at optical wavelengths.

\footnotetext[1]{Information available at http:// tempo.sourceforge.net/.}

\section{Data analysis}
We have employed the TEMPO\footnotemark  ~code, widely used  to model the pulse arrival times from pulsars, to construct a timing model of TVLM 513 radio flares and optical brightness variations based on the measurements listed above. In the process of modeling the possible deterministic changes of the flaring period with TEMPO, the number of flares received over a time interval $t=t_B-t_0$ between some initial epoch $t_0$ and the barycentric TOA, $t_B$, represents the measure of the accumulated phase $\phi$, which can be expressed in terms of the frequency $\nu=1/P$ and its time derivatives as a Taylor series:
\begin{equation} \phi~=~\phi_0~+~\nu~t~+~{1\over 2}\dot\nu~t^2~+~{1\over 6}
            \ddot\nu~t^3~+~\cdots .\end{equation} 
The parameters of the timing model are then determined as corrections to their initial values computed by means of a linearized least-squares fit of the model to the flare arrival times (see, for example, Lorimer \& Kramer 2005, for more details).  

\begin{figure}
\centering
\includegraphics[angle=0,width=0.99\textwidth]{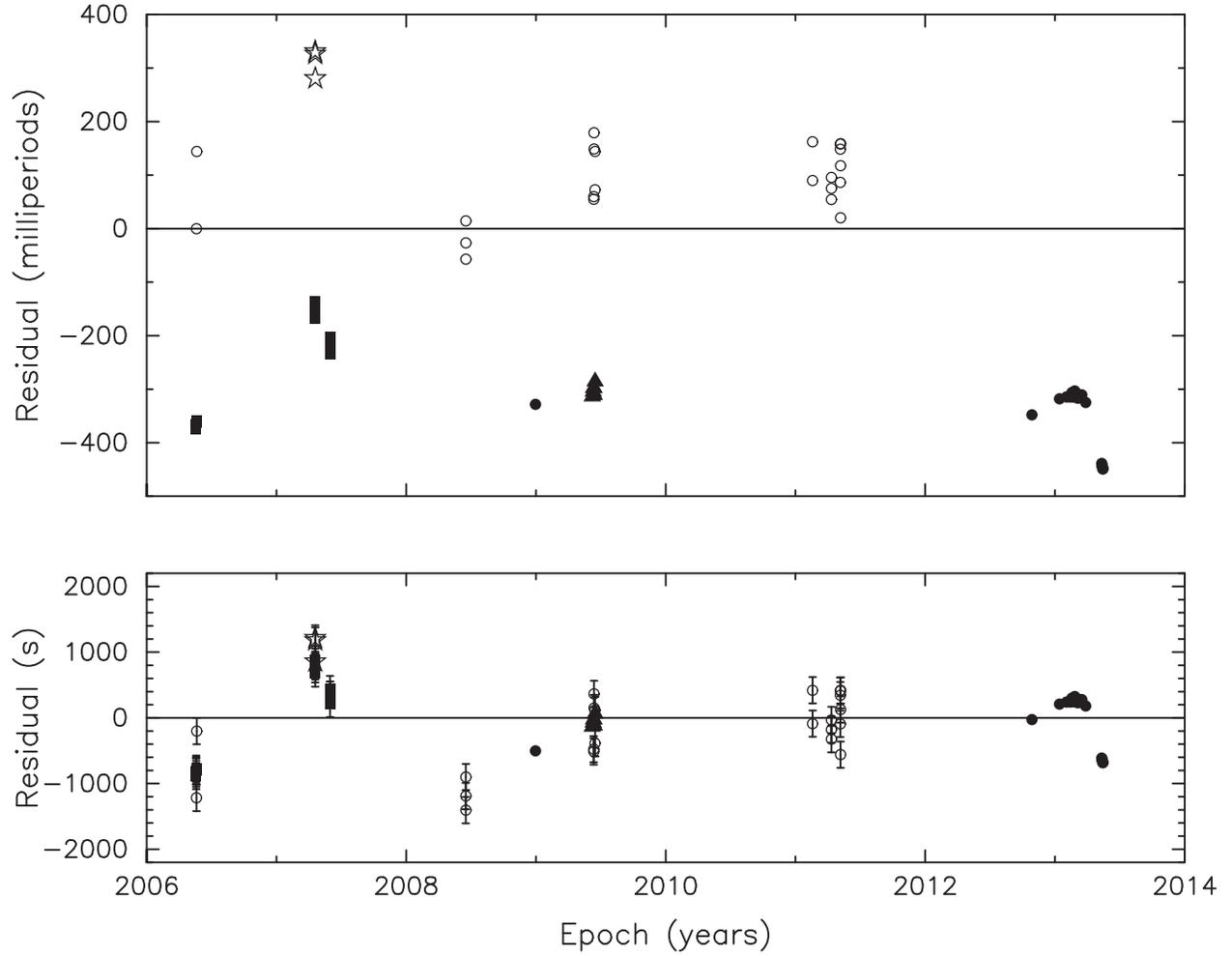}
\caption{\small Post-fit residuals from the modeling of TVLM 513 radio flare and optical brightness peak arrival times. (Top) A fit for phase only with the period fixed at 7054.49 s (Harding et al. 2013). (Bottom) A fit for phase, period, and the offset between the radio and the optical data. The open circles and stars denote the optical data taken from Harding et al. (2013) and Berger et al. (2008), respectively. The filled squares and triangles are for the respective radio measurements by Hallinan et al. (2006), Hallinan et al. (2009), and Doyle et al. (2010). The filled circles mark the radio flare arrival times derived from the Arecibo observations described in this paper. }
\label{fig2}
\end{figure}

Our model of the data consisting of the measured TOAs of the peaks of the optical light curves, and the TOAs of the radio flares between 2006 and 2013 is shown in Fig. \ref{fig2}. The model includes as free parameters the initial phase, $\phi_0$, the flaring period, $P$, and any possible phase offsets between the optical and the radio data. The initial guess value of the period of 7054.49$\pm$0.18 s has been taken from Harding et al. (2013). As this period has been derived by phase connecting the TVLM 513 brightness variations over a five-year time baseline that overlaps with ours, it is unlikely that our timing solution would suffer from any phase ambiguities. The post-fit rms residual for this model amounts to 567 s and is dominated by the slow phase variations caused by long-term changes of the flaring period and by the estimation errors of the TOAs of the broad peaks in the sine wave - like optical brightness variations. It is clear from Fig. \ref{fig2} that the residual phase variations caused by changes in the apparent flaring period are significantly larger than errors in the TOA measurements of both the radio flares and the optical brightness variations. The astrometric parameters of TVLM 513 (Table 1), which are also part of the timing model, have been taken from Forbrich et al. (2013). As the object's astrometric position is much too accurate for its errors to have any measurable effect on our flare TOA determinations, given their limited precision, these parameters have been kept fixed in the modeling process. 

The best-fit barycentric period of TVLM 513 derived from our model is equal to 7054.468$\pm$0.007 s, which is well within the much larger error limits of the period given by Harding et al. (2013), based on the shorter time baseline of the optical data alone and a different method of data analysis. Doyle et al. (2010) have measured a period of 7082.39$\pm$0.07 s from the two $\sim$8 hr VLA observations of TVLM 513 spaced by about six weeks in 2007, using Bayesian oscillation parameter estimation analysis (Marsh et al. 2008) to improve the precision of the flaring period estimate. Our period measurement using TEMPO to process the same data yielded 7053.62$\pm$0.12 s, which is almost 30 seconds shorter than the period computed by Doyle et al., but it is close to its long-term average derived from our model and from Harding et al. (2013). A closer scrutiny of this discrepancy shows that assuming the period computed by Doyle et al. leads to a drift of the expected flare arrival time which accumulates to almost exactly two periods over the 42 days between the two VLA measurements. It is clearly this phase ambiguity which has led to the corresponding ambiguity in the flaring period determination by these authors.

It is also clear from our timing model of TVLM 513 that the radio flares arrive approximately one-half of the period earlier than the peaks of the I-band, $i^{'}$-band, and H$\alpha$ light curves. The best-fit value for this phase offset is 0.41$\pm$0.02 of the period. Finally, it is evident from Fig. \ref{fig2} that the phase variations of the radio flare and optical peak TOAs track each other very well, and that there is no visible long-term leftover trend in the post-fit phase residuals. The model parameters are summarized in Table 1.
 
\begin{figure}
\centering
\includegraphics[angle=0,width=0.99\textwidth]{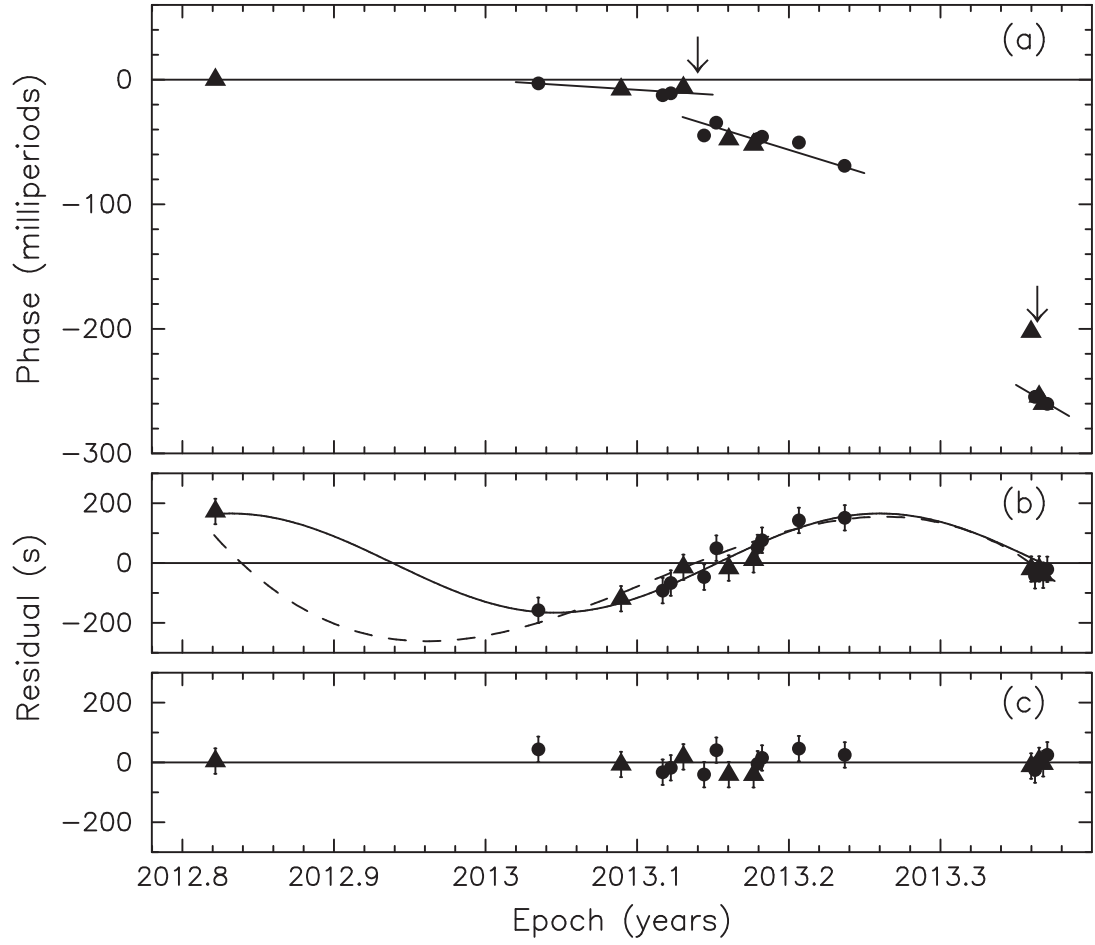}
\caption{\small Post-fit residuals from the modeling of TVLM 513 flare arrival times. (a) A fit for phase with a fixed period, and its time derivatives set to zero. Vertical arrows mark the two phase jumps. The sloping lines denote the best fits to subsets of data made for the purpose of period measurement. (b) Post-fit residuals from the model including phase, period, its first time derivative, and the two phase jumps. The solid and the dashed lines compare the fits of these residuals with the circular, 162-day orbit and the second derivative of the period, respectively. (c) A fit for phase, period, its first two time derivatives, and the phase jumps. Triangles mark the data points obtained from flares shown in Fig.1.}
\label{fig3}
\end{figure}

The results of a separate analysis of our most recent timing observations of TVLM 513, made between 2012 October and 2013 May with the Arecibo radio telescope, are shown in Fig. \ref{fig3} and listed in Table 1. These data are included in the model of Fig. \ref{fig2}, and consist of 19 flare TOA measurements spread over a $\sim$200 day time baseline. A relatively dense sampling of the flare TOAs makes this data set much better suited for a detailed, short term monitoring of the periodic behavior of the flares than any of the previously published observations. Evidently, the apparent flaring period of this UCD had been steadily decreasing over that period, with the accumulated phase shift reaching 25\% of the assumed constant period at the end of the observing campaign. In addition, one can see that the ``spin-up" process is not continuous, but it appears to get interrupted by jumps in phase after which the period becomes shorter than before the jump, as indicated by the apparently stepwise steepening of the slope of the phase residual. This timing behavior of the TVLM 513 flares can be satisfactorily described by the model that includes the period and its first two time derivatives, $P$=7054.7$\pm$0.1 s, $\dot P$=(-1.13$\pm0.07)\times$10$^{-7}$ s s$^{-1}$, and $\stackrel{..}{P}$=(-4.1$\pm0.7)\times$10$^{-14}$ s s$^{-2}$ at the reference epoch of MJD 56327.0, and the two jumps of the corresponding magnitudes of 200$\pm$40 s, and 320$\pm$50 s. It produces the 28 s rms post-fit residual, which agrees very well with the uncertainty of the flare TOA estimates discussed in Section 2 and is almost two orders of magnitude smaller than the accumulated residual phase drift caused by changes in the flaring period. This result also means that the flaring period of TVLM 513 has decreased by about 2 seconds in 200 days,  and that the rate of the period change has been accelerating.

As shown in Fig. \ref{fig3}, one can also fit the data with the model that includes $P$, $\dot P$ and a circular  Keplerian orbit to account for seemingly periodic post-fit residual variations left over from a fit for $P$ and $\dot P$ alone. This produces a 160-day, 0.42 AU orbit, which would require a minimum companion mass of 0.31 M$_{\odot}$ given the assumed 0.08 M$_{\odot}$ mass of TVLM 513. 
As such a stellar-mass, approximately M4-spectral type companion would have been easily detected by imaging observations conducted in the past (e.g. Close et al. 2003), and it has been ruled out by the VLA astrometry (Forbrich et el. 2013), explaining the timing behavior of TVLM 513 in terms of binary motion does not appear feasible.

\section{Discussion}
To our knowledge, the results described in this paper represent the first example of a timing analysis of the long-term, periodic behavior of a UCD, based on the arrival time measurements of its flares and optical brightness variations. The recent TVLM 513 rotation period measurement of 7054.49$\pm$0.18 s by Harding et al. (2013), derived from the optical data, also analyzed in this paper, has been obtained by applying a combination of the Lomb-Scargle periodogram (Lomb 1976; Scargle 1982) and the Phase Dispersion Minimization (Stellingwerf 1978) techniques, followed by a successive phase connection of the light curves over the 5 year time baseline. Our timing models fitted separately to the radio and the optical arrival time data produce nearly identical results of 7054.466$\pm$0.006 s and 7054.47$\pm$0.02 s, respectively, and the model based on both the radio and the optical data gives the period of 7054.468$\pm$0.007. This result improves the precision of the measurement by Harding et al. by a factor of $\sim$20, extends its temporal baseline from 5 to 7 years, and further strengthens the existing evidence for a long-term positional stability of the active region in TVLM 513. 

Our timing analysis also reveals that the radio and the optical TOAs of the brightness maxima from TVLM 513 have been maintaining the best-fit phase offset of 0.41$\pm$0.02 of the rotational period over the last 7 years Fig. \ref{fig2}. Berger et al. (2008) have analyzed the simultaneous radio and H$\alpha$ measurements of TVLM 513 taken in 2007 April and recognized that three of the observed radio bursts were approximately aligned with the minima of the H$\alpha$ light curves.  The apparent absence of the 1.96-hr periodicity in the radio data has led these authors to the conclusion that this phase coincidence was likely due to small number statistics. However, a subsequent reanalysis of the same radio data by Doyle et al. (2010) has shown that they do contain a series of consecutive, periodic flares. As shown in Fig. \ref{fig2}, our timing analysis of these flares along with the TOA measurements of the H$\alpha$ peaks from Berger et al. (2008) clearly confirms the existence of the $\sim$0.4 phase offset between these data sets.
  
This result strongly suggests that the radio flares and the optical brightness variations of TVLM 513 have a common origin, most likely associated with a large scale, approximately dipolar magnetic field that has been stable over at least the 7-year time baseline of the data analyzed here. A similar, multiwavelength analysis has been conducted for the UCD 2MASSW J0746425+200032 by Berger et al. (2009), who detected a 1/4 period phase shift between the periodic H$\alpha$ and radio peaks in the emission from this object. These authors have concluded that the observed phase shift can be explained in terms of a quadrupolar geometry of the magnetic field of the star, just like the shift by exactly 1/2 of the period would imply a magnetic dipole model, which seems to be approximately true for TVLM 513.

The above conclusions are further supported by the fact that both the radio and the optical data from TVLM 513, when analyzed separately over the respective 7-year and 5-year time baselines, produce the same period to within a precision of $\sim$20 ms. Interestingly, this seems not to be true for another periodically varying UCD, 2M J13142039+130011, in which case the radio and the optically determined periods are slightly different, possibly pointing to a differential rotation of the star as a cause of this discrepancy (McLean et al. 2011). 

\begin{figure}
\centering
\includegraphics[angle=0,width=0.99\textwidth]{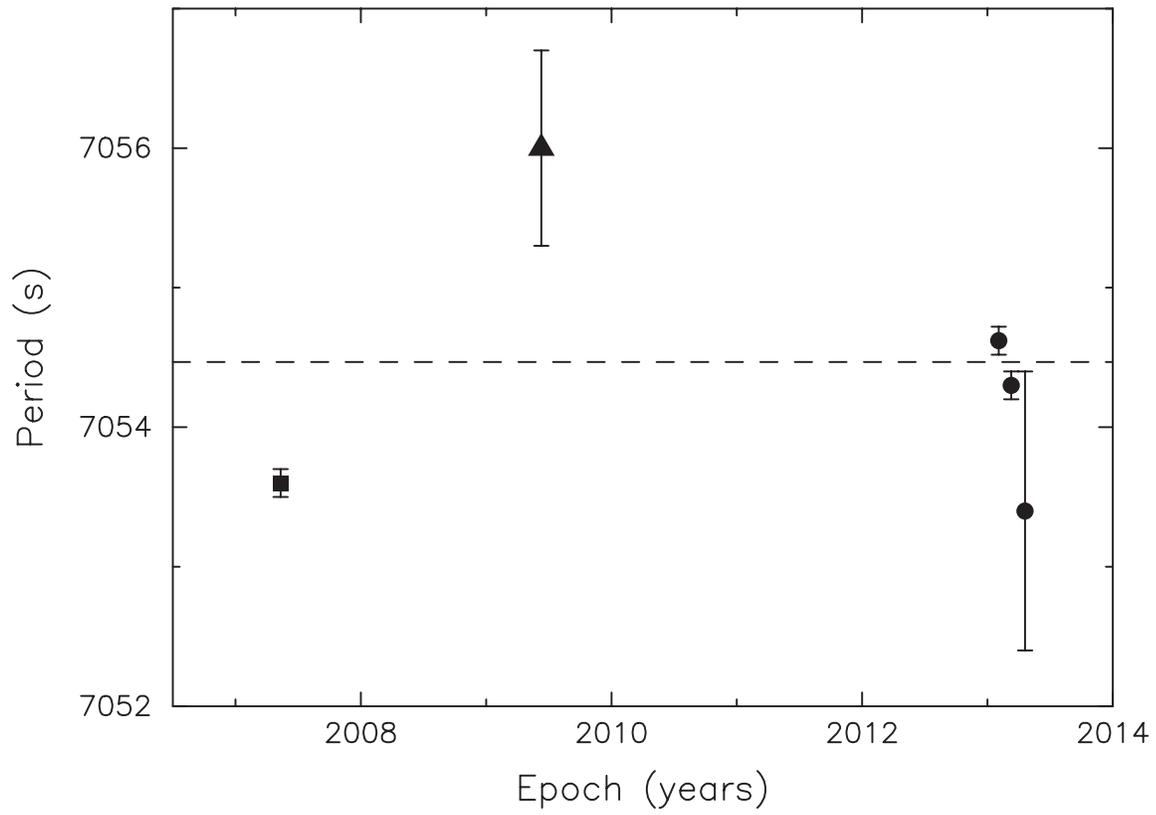}
\caption{\small Period variations of flares from TVLM 513. The filled square, triangle and circles represent period measurements based on the data from  Hallinan et al. (2009), Doyle et al. (2010), and the Arecibo measurements discussed in this paper, respectively. The horizontal, dashed line marks the best-fit period derived from the timing model of the entire set of arrival time measurements made between 2006 and 2013.}
\label{fig4}
\end{figure}

A correlated phase behavior of the TOAs of the radio flares and the brightness maxima of the light curves of TVLM 513 clearly shows that, between 2007 and 2013, the apparent period has been fluctuating about its best-fit value with a $\pm$1-2 s amplitude, and that the period changes must be caused by a mechanism that affects both kinds of emission in the same way. This is further demonstrated in Fig. \ref{fig4}, which shows period measurements using TEMPO fits to the radio data subsets spanning approximately week-to-month time baselines.

The timing model of the well-sampled set of radio flares observed at Arecibo in late 2012 - early 2013 and presented in Section 3 suggests that, in fact, the flaring period may have been changing in a highly systematic manner over the entire 2006-2013 time baseline covered by the data analyzed in this paper.  It is tempting to offer an explanation of this short-term timing behavior of TVLM 513 in terms of differential rotation (DR) of the star. DR plays an important role in the generation of stellar magnetic fields, and, for stars with radiative cores, it has been commonly observed (e.g. Barnes et al. 2005; Reinhold et al. 2013), and reasonably well understood (K\"uker \& R\"udiger 2011). In the case of fully convective stars, the relatively small number of available observations (Morin 2012) appear to support the theoretical prediction of a significant suppression of DR by large-scale magnetic fields (Browning 2008).

The observed stepwise shortening of the flaring period of TVLM 513 after phase jumps seen in the timing residuals (Fig. \ref{fig3}) suggests that the successive emission regions are created (or move into view) progressively closer to the equator and hence move faster with the star because of its DR, very much like the spots on the Sun (Hathaway 2010). Using the standard parametrization of the DR in terms of the surface shear, $\Delta\Omega$=$\Omega_{eq}-\Omega_{pol}$, where $\Omega_{eq}$ and $\Omega_{pol}$ are the respective values of angular spin at the equator and at the pole of the star, and the lapping time, t$_{lap}=2\pi/\Delta\Omega$ (K\"uker \& R\"udiger 2008), one gets $\Delta\Omega\ge$0.002 rad/day, and $t_{lap}\le$290 days, given the observed flaring period difference of $\ge$2 s, and the best-fit period of 7054.7 s. With all the uncertainties involved in such an approximate calculation, these estimates, when compared to the DR measurements for fully convective M-dwarfs (Morin et al. 2008a,b, 2010; Savanov \& Dmitrienko 2012), are entirely reasonable. 

\section{Conclusions}
In this paper, we demonstrate that the timing method, when applied to the periodic variations observed in the brightness of the ultracool dwarf, TVLM 513-46546, makes it possible to measure the period, detect its temporal evolution, and model it with the precision surpassing that of the techniques used in the past by an order of magnitude. This is because a modeling process that uses the TOAs of periodic brightness maxima deals with a cumulative error in phase, which makes it possible to measure very small deviations of model parameters from their correct values, given a sufficient timespan of the data (e.g. Lorimer \& Kramer 2005).

We show that the behavior of the apparent period of TVLM 513 measured over the 7-year span of the available radio and optical data is adequately described by the long-term, average value of 7054.47 s, representing the star's rapid  rotation, and the superimposed month-to-year timescale, $\sim$2 s variations, possibly due to the changing geometry of the emission region created by stellar magnetic activity. An upper limit to the fractional stability of the period, computed from the timing model of the entire dataset involving the period and its first time derivative, amounts to $\le$10$^{-5}$. Despite the fact that this value must still be biased by the observed shorter timescale period variations, it provides a useful quantification of the period stability of TVLM 513 and the related stability of its magnetic field deduced from earlier period measurements (Harding et al. 2013) and from the observed persistence of the star's radio emission (Hallinan et al. 2006; Hallinan et al. 2007; Doyle et al. 2010). 

Our analysis also reveals that the best-fit, long-term averaged period values computed separately from the optical and the radio measurements of TVLM 513 are practically the same. In addition, the TOA variations of brightness maxima observed in the two wavelength ranges are highly correlated but the radio peaks lead the optical ones by 0.4 of the period. Including this effect in the timing modeling shows that the phase offset has been maintained over the 7-year time baseline of the data to within 5\%. These new results point to a common origin of the observed radio and optical activity of TVLM 513, which must be powered by the star's approximately dipolar magnetic field. The deviation of the measured phase offset from its 0.5 value for the magnetic dipole is probably due to the field geometry not being exactly dipolar, or to the location of the active region with respect to the line of sight, or both.

Finally, our modeling of the high cadence timing observations of TVLM 513 made over 200 days in 2012 and 2013 clearly shows that the apparent period of radio flares from the star has been gradually becoming shorter, with the process being interrupted by small but easily detectable phase jumps in flare arrival times. We suggest that this phenomenon may be caused by a combination of a slow migration of the magnetically active region with the differential rotation of the star, in analogy to the well-known behavior of spots on the Sun and other stars. In fact, single measurements of stellar rotation scattered over time reveal period variations caused by a ``randomizing'' effect of a changing spot coverage of the star combined with its differential rotation (e.g. Henry et al. 1995; Barnes et al. 2005). Our work suggests that it should be possible to track the behavior of active regions in the magnetospheres of periodically flaring UCDs by means of multiwavelength timing of their brightness variations. If the active regions behave like stellar spots, it may be possible to use such measurements to monitor changes in the geometry of magnetically active areas and map their differential rotation in unprecedented detail. Overall, our results demonstrate that future timing measurements of TVLM 513 and other periodically varying ultracool dwarfs will definitely be of great value in the process of improving our understanding of the physics of these fascinating objects.

\section{Acknowledgements}

MR acknowledges support from the Center for Exoplanets and Habitable Worlds. The Center for Exoplanets and Habitable Worlds is supported by the Pennsylvania State University, and the Eberly College of Science. The Arecibo Observatory is operated by SRI International under a cooperative agreement with the National Science Foundation (AST-1100968), and in alliance with Ana G. M\'{e}ndez-Universidad Metropolitana, and the Universities Space Research Association.

\clearpage
\newpage

\clearpage
\newpage

\begin{table}
\begin{center}
\caption{Timing models for TVLM 513}
\vspace{0.1cm}
\begin{tabular}{ll}
\hline\hline
\multicolumn{2}{c}{\it Fixed parameters\,$^a$}\\
\hline
Right ascension, $\alpha$ (J2000) & $15^h 01^m 08\fs
157$ \\
Declination, $\delta$ (J2000)     & $22\degr 50\arcmin 01\,\farcs 425$\\
Proper motion, $\mu_{\alpha}$ (mas\,yr$^{-1}$)   & $-42.56$ \\
Proper motion, $\mu_{\delta}$   (mas\,yr$^{-1}$)   & $-65.47$ \\
Parallax, $\pi$ (mas) & $92.92$\\
Epoch (MJD)                              & $56188.65$ \\
\hline
\multicolumn{2}{c}{\it Best-fit parameters for the combined optical and radio data, 2006-2013}\\
\hline
Flare period, $P$ (s)                     & $7054.468\pm 0.007$\\
Phase offset of optical maxima and radio flares (s) & $2922 \pm 151$\\
Number of arrival times                        & 64\\
Post-fit RMS (s)           & 567.6        \\
\hline
\multicolumn{2}{c}{\it Best-fit parameters for the Arecibo radio data, 2012-2013}\\
\hline
Flare period, $P$ (s)                     & $7054.7\pm 0.1$\\
Period derivative, $\dot P$ (s\,s$^{-1}$) & (-1.13$\pm0.07)\times10^{-7}$\\
Period second derivative, $\ddot P$ (s\,s$^{-2})$ & $(-4.1\pm0.7)\times10^{-14}$\\
Number of arrival times                        & 19\\
Post-fit RMS (s)           & 28.0        \\
\noalign{\smallskip}\hline
\end{tabular}
\end{center}
{\small $^a$Astrometric parameters adopted from Forbrich et al. (2013).}\newline
\end{table}

\end{document}